\documentclass[aps,prl,twocolumn,showpacs,superscriptaddress]{revtex4-1}

\usepackage[final]{graphicx}
\usepackage{amsmath}
\usepackage{verbatim}
\usepackage{color}
\usepackage{ulem}
\usepackage[colorlinks,
linkcolor=blue,
citecolor=blue,
urlcolor=blue]{hyperref}

\begin{document}

\preprint{}

\title{Harmonics Effect on Ion-Bulk Waves in CH Plasmas}

\author{Q. S. Feng} 
\affiliation{HEDPS, Center for
	Applied Physics and Technology, Peking University, Beijing 100871, China}

\author{C. Y. Zheng} \email{zheng\_chunyang@iapcm.ac.cn}

\affiliation{HEDPS, Center for
	Applied Physics and Technology, Peking University, Beijing 100871, China}
\affiliation{Institute of Applied Physics and Computational
	Mathematics, Beijing, 100094, China}
\affiliation{Collaborative Innovation Center of IFSA (CICIFSA) , Shanghai Jiao Tong University, Shanghai, 200240, China}

\author{Z. J. Liu} 
\affiliation{HEDPS, Center for
	Applied Physics and Technology, Peking University, Beijing 100871, China}
\affiliation{Institute of Applied Physics and Computational
	Mathematics, Beijing, 100094, China}

\author{L. H. Cao} 
\affiliation{HEDPS, Center for
	Applied Physics and Technology, Peking University, Beijing 100871, China}
\affiliation{Institute of Applied Physics and Computational
	Mathematics, Beijing, 100094, China}
\affiliation{Collaborative Innovation Center of IFSA (CICIFSA) , Shanghai Jiao Tong University, Shanghai, 200240, China}

\author{C. Z. Xiao}
\affiliation{HEDPS, Center for
	Applied Physics and Technology, Peking University, Beijing 100871, China}

\author{Q. Wang}
\affiliation{HEDPS, Center for
	Applied Physics and Technology, Peking University, Beijing 100871, China}

\author{X. T. He} \email{xthe@iapcm.ac.cn}
\affiliation{HEDPS, Center for
	Applied Physics and Technology, Peking University, Beijing 100871, China}
\affiliation{Institute of Applied Physics and Computational
	Mathematics, Beijing, 100094, China}
\affiliation{Collaborative Innovation Center of IFSA (CICIFSA) , Shanghai Jiao Tong University, Shanghai, 200240, China}


\date{\today}

\begin{abstract}
	The harmonics effect on ion-bulk (IBk) waves has been researched by Vlasov simulation. The condition of excitation of a large-amplitude IBk waves is given to explain the phenomenon of strong short-wavelength electrostatic activity in solar wind. When $k$ is much lower than $k_{lor}/2$ ($k_{lor}$ is the wave number at loss-of-resonance point), the IBk waves will not be excited to a large amplitude, because a large part of energy will be spread to harmonics. The nature of nonlinear IBk waves in the condition of $k<k_{lor}/2$ is undamped Bernstein-Greene-Kruskal-like waves with harmonics superposition. Only when the wave number $k$ of IBk waves satisfies $k_{lor}/2\lesssim k\leq k_{lor}$, can a large-amplitude and mono-frequency IBk wave be excited. These results give a guidance for a novel scattering mechanism related to IBk waves in the field of laser plasma interaction.

\end{abstract}


\maketitle


The problem of understanding the role of kinetic effects and also fluid effects on the low frequency dynamics of collisionless plasmas, like solar wind, is nowadays of key significance in space physics. In the 1970s, several observations \cite{Gurnett_1977_1978_1979JGR} have shown that the most intense ion-acoustic waves (IAWs) usually occur in the low-velocity regions, i.e., in the large-wave-number regions. This Letter will give the corresponding analyses and the evidence consistent to these observations. However, in their analyses \cite{Gurnett_1977_1978_1979JGR}, only the short-wavelength IAWs were considered as the main electrostatic activity in solar wind. Recently, Valentini $et \ al.$ found that a novel branch of electrostatic kinetic waves was generated due to particles trapping in the development of the turbulent spectra \cite{Valentini_2008PRL_2009PRL}. The novel branch with a lower frequency than the IAWs and with a phase velocity $v_{\phi}$ close to the ion thermal velocity $v_{ti}$ is named ion-bulk (IBk) waves by Valentini \cite{Valentini_2011PRL}. Therefore, the nature of lower frequency waves with $v_{\phi}\simeq v_{ti}$ is not understood due to a strong Landau damping, which is similar to the electron-acoustic waves (EAWs) \cite{Valentini_2006POP}. By adding a external driving electric field (driver), after several bounce time of particles, the Landau damping of IBk waves will be decreased and be nearly zero \cite{Morales,(3),Feng_2016POP} due to particles trapping.
However, the energy was stored in the nonlinear IBk waves only when the wave number $k$ was larger than a special value \cite{Valentini_2008PRL_2009PRL}, which has not been explained and calls for a clear interpretation. This Letter will give a clear and detailed interpretation of these phenomena from the view of harmonics effects.

On the other hand, the research of IBk waves will give a guidance to a new scattering mechanism in the field of laser plasmas interaction (LPI) \cite{Lindl_2004POP,Glenzer_2010Science} related to inertial confinement fusion (ICF) \cite{Glenzer_2007Nature, He_2016POP}. As we know, laser scattering mainly includes stimulated Brillouin scattering (SBS) \cite{Liu_2011POP} from IAWs, stimulated Raman scattering (SRS) \cite{Xiao_2015POP} from Langmuir waves (LWs). In addition, Montgomery $et \ al.$ reported observation of a novel stimulated electron-acoustic-wave scattering (SEAS) \cite{Montgomery_2001PRL} related to electron-acoustic waves (EAWs) \cite{Xiao_2014POP}. However, Montgomery $et \ al$. \cite{Montgomery_2001PRL} thought that the accessibility of SEAS was only with a limit of $k\lambda_{De}<k_{lor}\lambda_{De}$, where $k_{lor}$ is the wave number at loss-of-resonance point defined later. While the harmonics effect on IBk waves was not considered. Inspired by SEAS, coupling of the intense laser field to IBk waves will result in a novel scattering mechanism, called stimulated ion-bulk-wave scattering (SIBS) in this Letter. Since IBk waves with strong Landau damping can only exist by particles trapping, resulting in undamped Bernstein-Greene-Kruskal-like (BGK-like) waves \cite{BGK}; the IBk-waves amplitude is usually lower than the IAWs amplitude. Thus, the SIBS is usually weaker than SBS, and the competitions of IBk waves and the IAWs can limit the scattering level \cite{Liu_2012PPCF}. This Letter will give a parameter scope to excite large-amplitude IBk waves, thus giving a clear guidance to SIBS in the field of LPI.

In this Letter, we report the harmonics effect on IBk waves especially when the wave number $k$ is lower than $k_{lor}/2$. The IBk waves with strong Landau damping can be excited by particles trapping, which will flat the distribution function at the phase velocity and thus reduce the Landau damping. Only when $k_{lor}/2\lesssim k\leq k_{lor}$, can the large-amplitude and single-frequency IBk waves be excited. If the wave number $k$ of IBk waves is much lower than $k_{lor}/2$, the IBk waves will nearly not be excited to a large amplitude and the energy will be spread to harmonics. The nature of the IBk waves in the condition of $k<k_{lor}/2$ is BGK-like modes with a superposition of harmonics. These results can not only give a clear interpretation of the strong short-wavelength electrostatic activity in solar wind, but also give a convincing guidance of a new scattering mechanism related to IBk waves in the field of LPI.


 \begin{figure}[!tp]
 	\includegraphics[width=1.0\columnwidth]{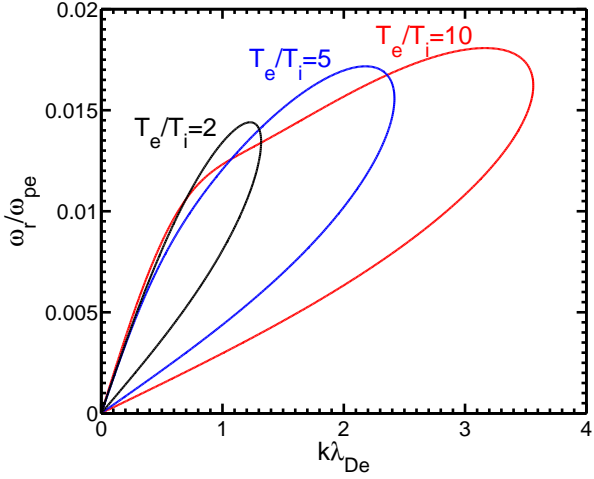}
 	
 	\caption{\label{Fig:Dispersion}(Color online) The dispersion relation of the ion-acoustic waves (the upper branches) and the ion-bulk waves (the lower branches) in the condition of $T_e/T_i=2, 5, 10$.}
 \end{figure}

If the duration time of the driver is several times of ion bounce time $\tau_{bi}=2\pi/\sqrt{kq_iE/m_i}$ ($q_i,\ m_i$ are the charge and mass of ion $i$, respectively; and $k,\ E$ are the wave number and electric amplitude of electrostatic waves, respectively), the Landau damping will be turned off, i.e.,  $Im(\omega)\approx0$ \cite{Morales,(3)}. Assuming that the electron temperature equals $T_e$ and the same temperature of all ions equals $T_i$, the dispersion relation of the infinitesimal amplitude IBk waves in non-magnetized, homogeneous plasmas consisting of multi-ion species is given by \cite{Feng_2016POP}
\begin{equation}
\label{Eq:Nonlinear}
Re(\epsilon_L(Re(\omega), k))=0,
\end{equation}
where $Re(\omega)$ is the real part of frequency of the infinitesimal amplitude nonlinear mode by taking $Im(\omega)=0$. The width of the plateau, where $\partial f_0/\partial v=0$, is infinitesimal, thus the distribution of all species is nearly a Maxwellian distribution. $\epsilon_L$ is defined as
\begin{equation}
\epsilon_L(\omega,k)=1+\sum_j \frac{1}{(k\lambda_{Dj})^2}(1+\xi_jZ(\xi_j)),
\end{equation}
where $Z(\xi_j)$ is the dispersion function

\begin{equation}
 Z(\xi_j)=1/\sqrt{\pi}\int_{-\infty}^{+\infty}e^{-v^2}/(v-\xi_j)dv,
\end{equation}
 and $\xi_j=\omega/(\sqrt{2}kv_{tj})$ is complex; $v_{tj}=\sqrt{T_j/m_j}$ is the thermal velocity of specie $j$; and $j$ represents electrons, H ions or C ions. $\lambda_{Dj}=\sqrt{T_j/4\pi n_jZ_j^2e^2}$ is the Debye length; and $T_j, n_j, m_j, Z_j$ are the temperature, density, mass, and charge number of specie $j$, respectively. 

By solving Eq. (\ref{Eq:Nonlinear}), one can get the dispersion relation of the IAWs (the upper branches) and the IBk waves (the lower branches) in different $T_e/T_i$ as shown in Fig. \ref{Fig:Dispersion}. We define the point, at which the minimum phase velocity of the ion-acoustic (IA) branch coincides with the maximum phase velocity of the ion-bulk (IB) branch, as a loss-of-resonance (LOR) point \cite{Rose_2001POP}. We will denote the wave number at the LOR point by $k_{lor}$. Beyond $k_{lor}$, no resonance is possible. As $T_e/T_i$ decreases, $k_{lor}$ will decrease.

To excite IBk waves, one dimension in space and velocity (1D1V) Vlasov-poisson code \cite{Liu_2009POP,Liu_2009POP_1} is used, and all of the particles including electrons and ions are considered as kinetic particles. We split the time-stepping operator of Vlasov equation into free-streaming in $x$ and motion in $v_x$, thus getting the advection equations \cite{7_5-1976JCP,7-2004CPC}. Then, the advection equations are solved by a third order Van Leer (VL3) scheme \cite{VL3, 10-2006POP} or the piecewise parabolic method (PPM) \cite{PPM}. To solve the particles behavior in CH plasmas (1:1 mixed), the spatial scale is $[0, L_x]$ (where $L_x=2\pi/k$) discretized with $N_x=64$ grid points, and the velocity scale is [$-v_{max}, v_{max}$] (where $v_{max}=8v_{tj}$) with $N_v=128$ grid points. The time step is $dt=0.02\omega_{pe}^{-1}$ and the periodic boundary condition in the spatial domain is used. The driving electric field (driver) form is  
\begin{equation}
\tilde{E}_d(x,t)=\frac{\tilde{E}_d^{max}}{1+[(t-t_0)/\Delta\tau]^n}\text{sin}(k_dx-\omega_dt),
\end{equation}
 in this Letter, $\tilde{E}_d^{max}=1\times10^{-3}$, $t_0=1\times10^5\omega_{pe}^{-1}$, $\Delta\tau=5\times10^4\omega_{pe}^{-1}$, $n=10$, and $[k_d, \omega_d]$ is the wave number and the frequency of the driver, separately.

\begin{figure}[!tp]
	
	\includegraphics[width=1.0\columnwidth]{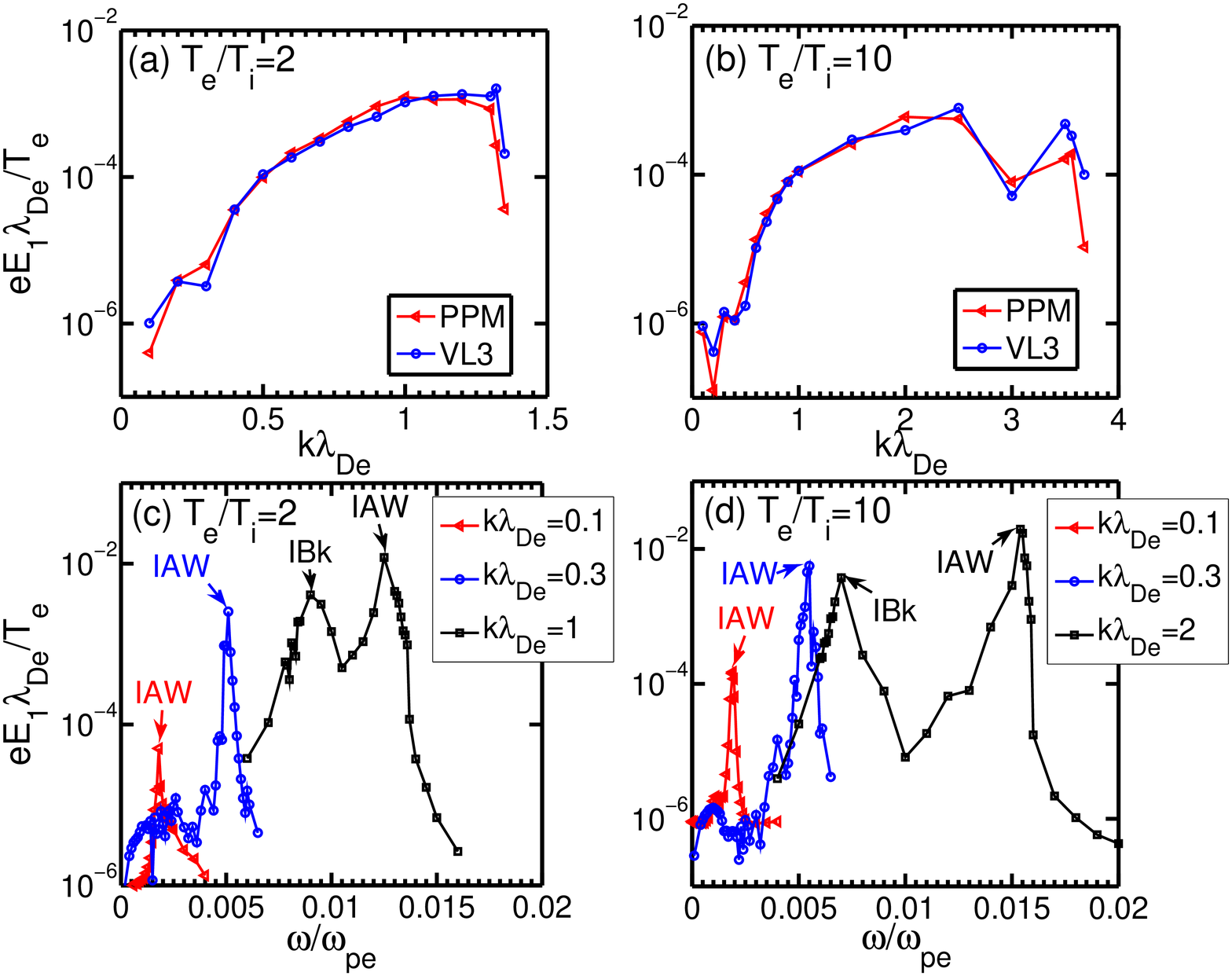}
	
	\caption	{\label{Fig:E1_k_E1_w}(Color online) The fundamental mode amplitude varies with the wave number $k\lambda_{De}$ in the condition of (a) $T_e/T_i=2$ and (b) $T_e/T_i=10$. The result of PPM algorithm (the red line) is consistent to that of the VL3 algorithm (the blue line). The resonance curves in the condition of (c) $T_e/T_i=2$ and (d) $T_e/T_i=10$.}
\end{figure}

\begin{figure}[!tp]
	
	\includegraphics[width=1.0\columnwidth]{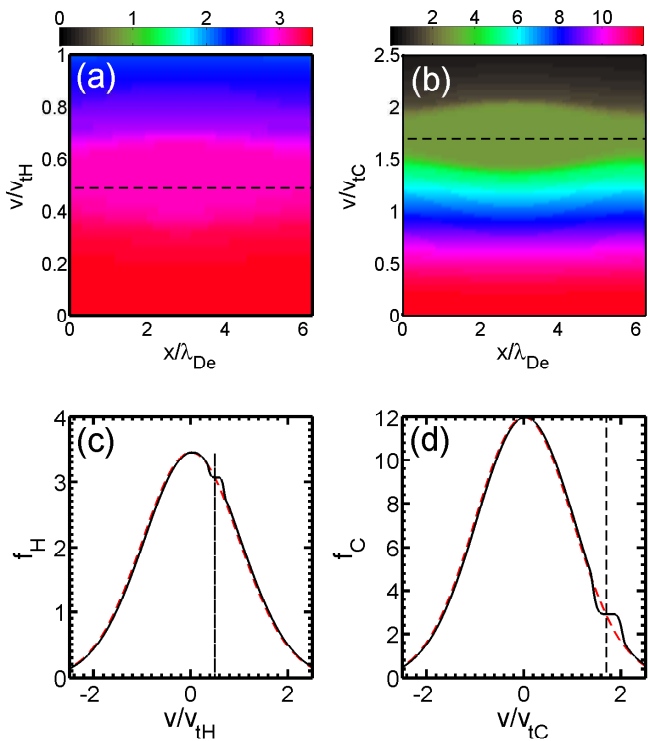}
	
	\caption	{\label{Fig: Phase pictures}(Color online) The phase pictures of (a) H ions and (b) C ions in the condition of $T_e/T_i=2, k\lambda_{De}=1.0$. Panels (c) and (d) are the corresponding distribution of H ions and C ions. Where the red dashed lines are the initial Maxwellian distributions and the black solid lines are the distributions at the time $\omega_{pe}t=3.2\times10^5$ (after the driver is off). The black dashed lines are the phase velocities of the IBk waves.}
\end{figure}

\begin{figure*}[!tp]
	
	\includegraphics[width=2\columnwidth]{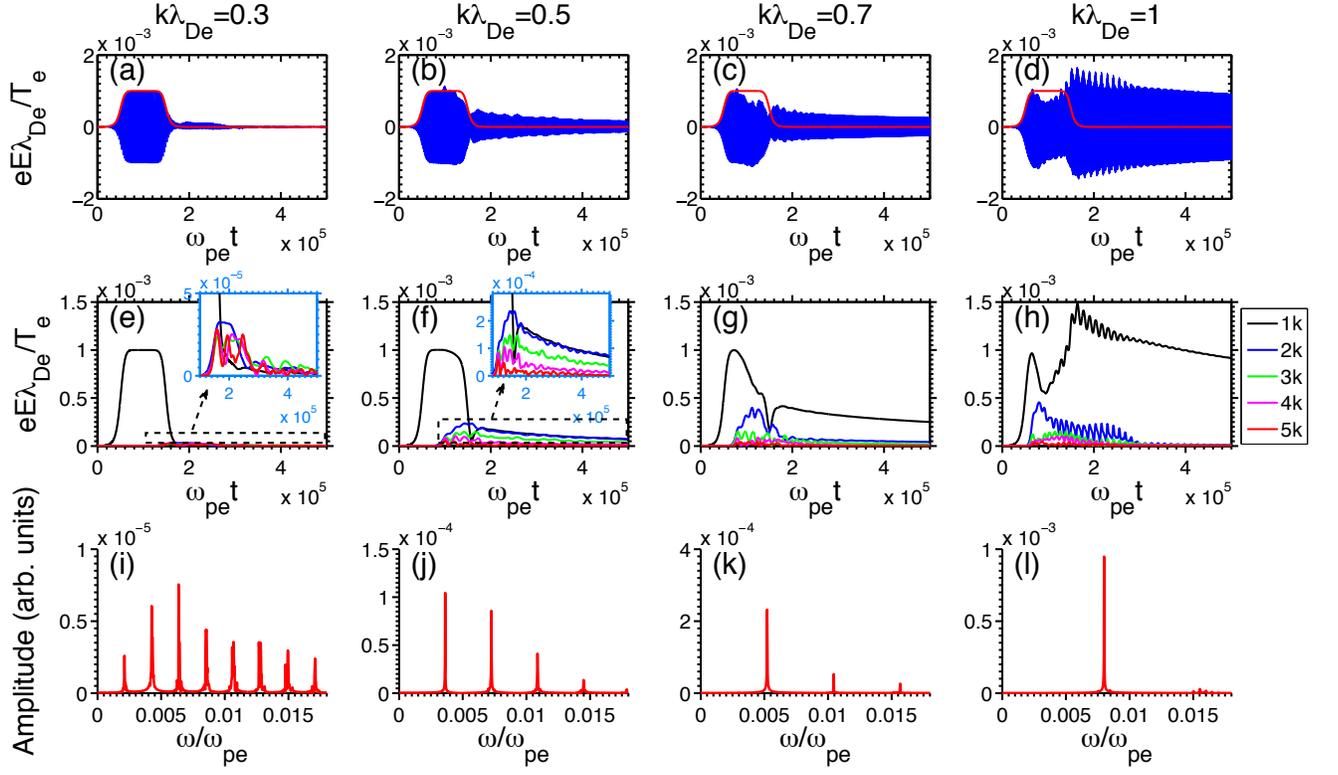}
	
	\caption	{\label{Fig4}(Color online) (a)-(d): The time evolution of the total electric field of the IBk wave (blue line) and the driver envelope (red line). (e)-(f): The amplitude of the top five harmonics evolve with time. Where $1k, 2k, 3k, 4k, 5k$ represent for the first, second, third, fourth and fifth harmonics. (i)-(l): The frequency spectra of the electric field $E(x,t)$ at $x_0=5\lambda_{De}$ and $\omega_{pe}t\in[2\times10^{5}, 5\times10^{5}]$. The condition of $T_e/T_i=2$ is fixed and the wave number $k\lambda_{De}$ varies.}
\end{figure*}

Figure \ref{Fig:E1_k_E1_w} shows the response of the IBk wave in different $k\lambda_{De}$ (Figs. \ref{Fig:E1_k_E1_w}(a) and \ref{Fig:E1_k_E1_w}(b)) and the resonance curves in the condition of different $k\lambda_{De}$ (Fig. \ref{Fig:E1_k_E1_w}(c) and \ref{Fig:E1_k_E1_w}(d)). For Figs. \ref{Fig:E1_k_E1_w}(a) and \ref{Fig:E1_k_E1_w}(b), the wave number and the frequency of the driver [$k_d, \omega_d$] take the values of the infinitesimal-amplitude IBk wave as shown in Fig. \ref{Fig:Dispersion}. And VL3 method and PPM are used to verify the consistent results. The VL3 method is more efficient than PPM, thus VL3 method is taken as the simulation algorithm in the following. As shown in Figs. \ref{Fig:E1_k_E1_w}(a) and \ref{Fig:E1_k_E1_w}(b), the fundamental mode amplitude generally increases as $k\lambda_{De}$ increases, but decreases rapidly when $k\lambda_{De}$ is beyond $k_{lor}$. This results give us a parameter scale 
\begin{equation}
k_{lor}/2\lesssim k\leq k_{lor}
\end{equation}
 to excite a large amplitude IBk wave. 
As we know, there is a nonlinear frequency shift (NFS) of the large amplitude BGK modes \cite{BGK} due to particles trapping and harmonic generation, such as the nonlinear IAWs \cite{Feng_2016POP, Berger_2013POP, Chapman_2013PRL,  Feng_2016PRE}, Langmuir waves (LWs) \cite{Rose_2001POP}, and nonlinear electron acoustic waves (EAWs) \cite{Valentini_2006POP}. To obtain the resonance curves as shown in Figs. \ref{Fig:E1_k_E1_w}(c) and \ref{Fig:E1_k_E1_w}(d), the wave number $k\lambda_{De}$ is fixed, and the driver is then swept in frequency from the frequency of the IBk waves to the IAWs.
When the IBk waves or the IAWs is excited to a large amplitude, the frequency of the maximum response will slightly deviate from the frequency of the corresponding infinitesimal-amplitude modes calculated by Eq. (\ref{Eq:Nonlinear}) due to NFS.  Whether the IBk waves can be excited to a large amplitude is not only related to the frequency of the driver $\omega_d$, but also decided by $k\lambda_{De}$. As shown in Figs. \ref{Fig:E1_k_E1_w}(c) and \ref{Fig:E1_k_E1_w}(d), when $k$ is much lower than $k_{lor}/2$, such as $k\lambda_{De}=0.1, 0.3$, the IBk waves will nearly not be excited by the driver. However, when $k_{lor}/2\lesssim k\leq k_{lor}$, the IBk waves can be excited to a large amplitude, which is even comparable to that of IAWs. These results give the proof of the strong kinetic effects of short-wavelength (large wave number) longitudinal IAWs and IBk waves on the low frequency dynamics of collisionless plasmas in solar wind \cite{Valentini_2008PRL_2009PRL,Gurnett_1977_1978_1979JGR}. On the other hand, the wave number $k\lambda_{De}$ of longitudinal waves is related to the electrons temperature $T_e$ and the electrons density $n_e$ in the field of LPI, i.e., 
\begin{equation}
k\lambda_{De}\simeq 2\frac{v_{te}}{c}\sqrt{n_c/n_e-1}, 
\end{equation}
where $v_{te}=\sqrt{T_e/m_e}$ is the electrons thermal velocity, $n_c$ is the critical density, and $c$ is the light speed. Thus, if the electrons temperature and the electrons density satisfy a condition that the wave number of IBk modes is in the scope of $k_{lor}/2<k<k_{lor}$, the strong SIBS will occur.
These provide the guidance for a novel scattering mechanism related to IBk waves in ICF.

For the phase velocity of the IBk waves is near the thermal velocity of ions, the Landau damping of IBk waves is very strong. Thus, the IBk waves are difficult to be excited and their amplitude is in general lower than the IAWs amplitude as shown in Figs. \ref{Fig:E1_k_E1_w} (c) and \ref{Fig:E1_k_E1_w}(d). However, when the duration time of the driver is long enough, the particles will be trapped by the electric field excited by the driver. The Landau damping will decrease and even be turned off \cite{Morales}. This process allows phase locking of the IBk waves to the driver, thus feedback is not required to maintain resonance \cite{Chapman_2013PRL}. As shown in Figs. \ref{Fig: Phase pictures}(a) and \ref{Fig: Phase pictures}(b), the particles are trapped near the phase velocity of IBk waves. And the distribution of the particles will flat at the phase velocity of the IBk waves as shown in Figs. \ref{Fig: Phase pictures}(c) and \ref{Fig: Phase pictures}(d). If the width of the plateau, where $\partial{f_0}/\partial v=0$, is infinitesimal, the assumption of Maxwellian distribution will be satisfied and the Landau damping will be turned off. Thus, the IBk waves will exist as the form of the dispersion relation calculated by Eq. (\ref{Eq:Nonlinear}) due to particles trapping.

To clarify why the IBk waves can not be excited to a large amplitude when $k<k_{lor}/2$, the case of $T_e/T_i=2$ is chosen as an example. As shown in Fig. \ref{Fig4}, in the case of $k\lambda_{De}=0.3$, which is much lower than $k_{lor}\lambda_{De}/2\simeq0.66$, the IBk wave can nearly not be excited after the driver is off (Fig. \ref{Fig4}(a)). The nature of this phenomenon is that the harmonics such as the second and higher-order harmonics will be excited resonantly when $k<k_{lor}/2$. The related researches about resonantly excited nonlinear IAWs \cite{Cohen_1997POP} and wave-wave interactions of nonlinear EAWs \cite{Xiao_2014POP} have been made. Inspired by these research, the electric field amplitude of the second harmonic $eE_2\lambda_{De}/T_e$ (denoted as $\tilde{E}_2$) scale as $\delta n_i^2/\Delta_{2k}$, i.e., 
\begin{equation}
\label{Eq: E2}\tilde{E}_2\sim \delta n_i^2/\Delta_{2k},
\end{equation}
where $\delta n_i$ is the density perturbations of ions, and $\Delta_{2k}$ is the frequency mismatch between twice the fundamental frequency $2\omega_{k}$ and the second-harmonic resonance frequency $\omega_{2k}$ calculated by the dispersion relation for IBk waves (Fig. \ref{Fig:Dispersion}) at $2k$, i.e., $\Delta_{2k}=|2\omega_k-\omega_{2k}|$. And the higher-order harmonics will obey the similar physical laws as Eq. (\ref{Eq: E2}):
\begin{equation}
\tilde{E}_n\sim \delta n_i^2/\Delta_{nk},
\end{equation}
where $n$ is the order of harmonics and $\Delta_{nk}=|n\omega_k-\omega_{nk}|$. 
If $k$ is much lower than $k_{lor}/2$, such as $k\lambda_{De}=0.3$, the second and higher-order harmonics will be excited resonantly. For a large energy is transferred and spread to the harmonics (many orders harmonics), there will not be a large amplitude of the IBk electric field (Fig. \ref{Fig4}(a)) when the driver is off. If $k$ is not much lower than $k_{lor}/2$, such as $k\lambda_{De}=0.5$, the second harmonic will be excited to a large amplitude comparable with the fundamental mode (or the first harmonic labelled as \textquotedblleft 1k\textquotedblright). In this case, the electric field of IBk waves (Fig. \ref{Fig4}(b)) are the superposition of the fundamental mode and the harmonics especially the second harmonic.
Since there is a LOR point, beyond $k_{lor}$, no resonance is possible. Thus, if $k_{lor}/2< k\le k_{lor}$, there will not be the resonance frequency of the second harmonic $\omega_{2k}$ ($2k>k_{lor}$), as a result, the second and higher-order harmonics will not be excited resonantly. And the energy of the driver will be mainly transferred to the fundamental mode, thus, the large-amplitude and single-frequency IBk waves will be excited. The cases of $k\lambda_{De}=0.7$ (Fig. \ref{Fig4}(c), \ref{Fig4}(g), \ref{Fig4}(e)) and $k\lambda_{De}=1$ (Fig. \ref{Fig4}(d), \ref{Fig4}(h), \ref{Fig4}(l)) show the results consistent with the analyses above. The harmonics behaviors of IAWs obey the same physical laws above as IBk waves, but the IAWs amplitude is generally larger than IBk-waves amplitude as shown in Figs. \ref{Fig:E1_k_E1_w}(c) and \ref{Fig:E1_k_E1_w}(d), since the Landau damping of IBk waves is larger than that of IAWs. Although this letter discusses the behavior of IBk waves in CH plasmas, the same physical laws are applicable to other multi-ion species plasmas such as $\text{He}_{1}\text{H}_{3}$ or H plasmas in solar wind. These results give a clear explanation of why the most intense IAWs usually occur in the low-velocity (or large-wave-number) regions \cite{Gurnett_1977_1978_1979JGR}, and also explain why a novel branch called IBk waves can exist in solar wind only when the wave number $k$ is large \cite{Valentini_2008PRL_2009PRL}.

In summary, we have reported the first clear evidence of the harmonics effect on IBk waves in multi-ion species plasmas. Only when the wave number satisfies the condition $k_{lor}/2\lesssim k\le k_{lor}$, can the large-amplitude and single-frequency IBk waves be excited. The nature of low-amplitude IBk waves in the condition of $k\lesssim k_{lor}/2$ is the BGK-like waves with a superposition of harmonics. If $k$ is much lower than $k_{lor}/2$, the energy will be transferred and spread to harmonics, thus the large-amplitude IBk waves will not be excited. These results give a clear interpretation of significant levels of short-wavelength (large-wave-number) longitudinal IAWs in several solar-wind observations \cite{Gurnett_1977_1978_1979JGR}, and also explain  why large-amplitude IBk waves exist in solar wind only when $k$ is large \cite{Valentini_2008PRL_2009PRL}. Furthermore, this letter give a prediction of a new scattering mechanism related to IBk waves in the field of laser plasma interaction (LPI), if the electrons temperature is high and the electrons density is low enough.\\

\begin{acknowledgments}
We are pleased to acknowledge useful discussions with K. Q. Pan. This research was supported by the National Natural Science Foundation of China (Grant Nos. 11575035, 11475030 and 11435011) and National Basic Research
Program of China (Grant No. 2013CB834101).
\end{acknowledgments}


\end{document}